%% file: arXiv_version.tex
\documentclass[journal=jacsat,manuscript=communication,layout=twocolumn]{achemso}

\usepackage{chemformula} 
\usepackage[T1]{fontenc} 

\usepackage{graphicx} 
\graphicspath{ {./images/} }

\usepackage[rightcaption]{sidecap}

\usepackage{wrapfig}
\usepackage{amssymb}
\mciteErrorOnUnknownfalse

\SectionsOn


\author{James Eills}
\affiliation{Institute for Bioengineering of Catalonia, Barcelona Institute of Science and Technology, 08028 Barcelona, Spain}
\alsoaffiliation{Helmholtz-Institut Mainz, GSI Helmholtzzentrum f{\"u}r Schwerionenforschung, 55128 Mainz, Germany}
\alsoaffiliation{Institute for Physics, Johannes Gutenberg-Universit{\"a}t  Mainz, 55099 Mainz, Germany}
\email{jeills@ibecbarcelona.eu}

\author{Rom\'{a}n Picazo-Frutos}
\affiliation{Helmholtz-Institut Mainz, GSI Helmholtzzentrum f{\"u}r Schwerionenforschung, 55128 Mainz, Germany}
\alsoaffiliation{Institute for Physics, Johannes Gutenberg-Universit{\"a}t  Mainz, 55099 Mainz, Germany}

\author{Oksana Bondar}
\affiliation{Department of Molecular Biotechnology and Health Sciences, Center of Molecular Imaging, University of Turin, 10126 Turin, Italy}

\author{Eleonora Cavallari}
\affiliation{Department of Molecular Biotechnology and Health Sciences, Center of Molecular Imaging, University of Turin, 10126 Turin, Italy}

\author{Carla Carrera}
\affiliation{Institute of Biostructures and Bioimaging, National Research Council of Italy, Turin, 10126 Italy}

\author{Sylwia J. Barker}
\affiliation{School of Chemistry, University of Southampton, SO17 1BJ Southampton, United Kingdom}

\author{Marcel Utz}
\affiliation{School of Chemistry, University of Southampton, SO17 1BJ Southampton, United Kingdom}

\author{Silvio Aime}
\affiliation{Department of Molecular Biotechnology and Health Sciences, Center of Molecular Imaging, University of Turin, 10126 Turin, Italy}

\author{Francesca Reineri}
\affiliation{Department of Molecular Biotechnology and Health Sciences, Center of Molecular Imaging, University of Turin, 10126 Turin, Italy}

\author{Dmitry Budker}
\affiliation{Helmholtz-Institut Mainz, GSI Helmholtzzentrum f{\"u}r Schwerionenforschung, 55128 Mainz, Germany}
\alsoaffiliation{Institute for Physics, Johannes Gutenberg-Universit{\"a}t  Mainz, 55099 Mainz, Germany}
\alsoaffiliation{Department of Physics, University of California at Berkeley, 94720 Berkeley, U.S.A.}

\author{John W. Blanchard}
\affiliation{Helmholtz-Institut Mainz, GSI Helmholtzzentrum f{\"u}r Schwerionenforschung, 55128 Mainz, Germany}
\alsoaffiliation{NVision Imaging Technologies GmbH, 89081 Ulm, Germany}

\title{Metabolic Reactions Studied by Zero- and Low-Field Nuclear Magnetic Resonance}

\begin{document} 

\begin{abstract}
State-of-the-art magnetic resonance imaging uses hyperpolarized molecules to track metabolism in vivo, but large superconducting magnets are required, and the strong magnetic fields largely preclude measurement in the presence of conductive materials and magnify problems of magnetic susceptibility inhomogeneity.
Operating at zero and low field circumvents these limitations, but until now has not been possible due to limited sensitivity.
We show that zero- and low-field nuclear magnetic resonance can be used for probing two important metabolic reactions: the conversion of hyperpolarized fumarate to malate and pyruvate to lactate.
This work paves the way to a heretofore unexplored class of biomedical imaging applications.
\end{abstract}

\section{Introduction}

Nuclear magnetic resonance (NMR) is among the most important spectroscopic techniques for chemical analysis and biochemical structure elucidation, and the imaging modality (MRI) is an invaluable diagnostic tool for non-invasive medical imaging. 
NMR has the potential to be even more widely applicable if the nuclear-spin polarization could be boosted beyond its thermal equilibrium value of $\approx 10^{-5}$ for samples at room temperature in experimentally achievable magnetic fields.
In recent years, nuclear-spin hyperpolarization techniques -- physical and chemical methods to increase NMR signals by 4-5 orders of magnitude\,\cite{kuhn2013hyperpolarization,nikolaou2015NMR} -- have enabled a new application: hyperpolarized metabolic imaging.\cite{wang2019hyperpolarized} Small molecules can be hyperpolarized and injected in vivo, and tracking their metabolism can yield diagnostic information about disease progression and treatment response.\cite{nelson2013metabolic,gallagher2009production}
Even with enhanced signals, conventional high-field magnetic resonance has some drawbacks: (1) high-resolution spectroscopy is generally only achievable in homogeneous samples; (2) the high-frequency signals are distorted by and do not readily penetrate conductive materials, due to the skin effect; (3) MRI scans are not advised for women in the early stages of pregnancy, or for patients with medical implants, and; (4) high-field magnets are bulky, heavy, and expensive to operate - they cannot be deployed in the field which hinders, for example, point-of-care diagnostics.

\begin{figure*}
\includegraphics[width=0.9\textwidth]{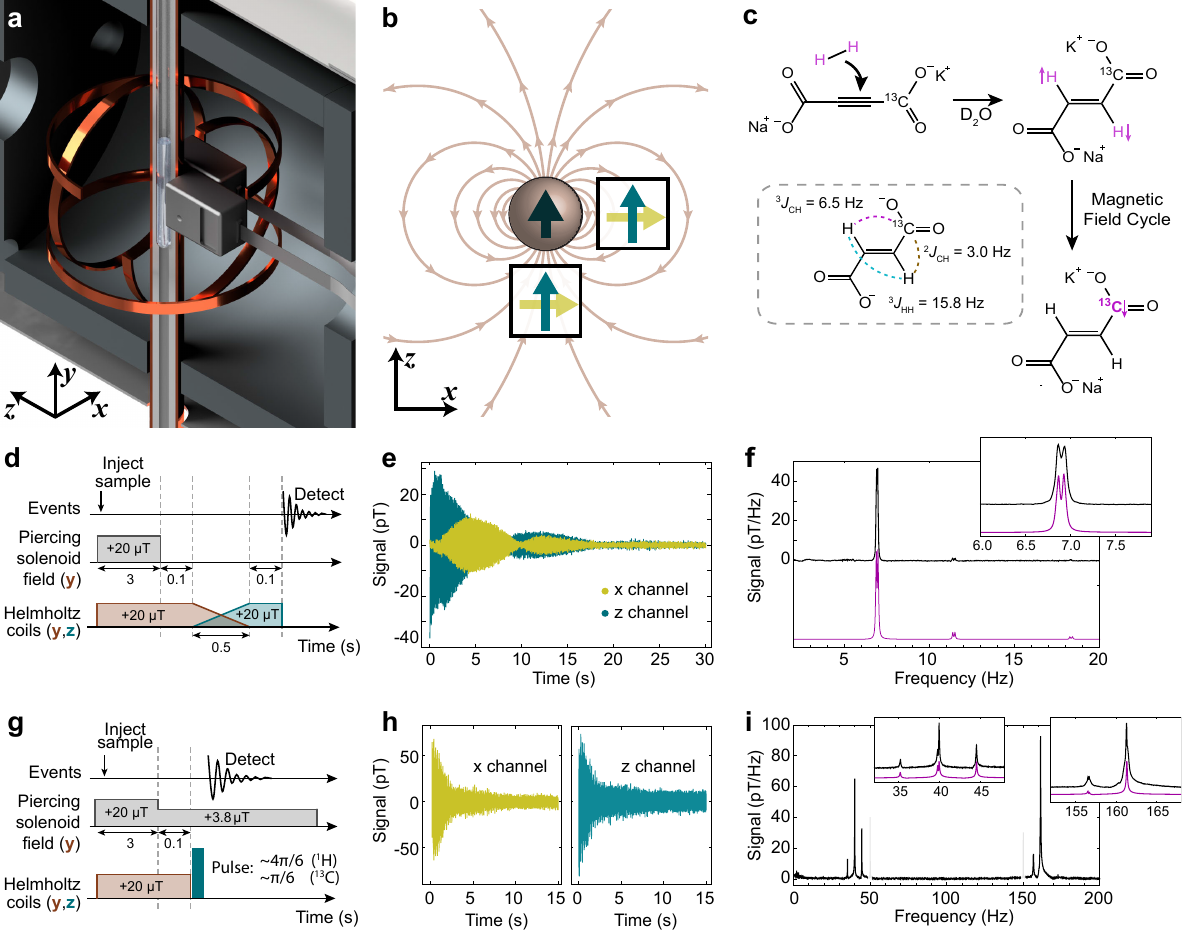}%
\caption{
\label{fig:Fig1} 
Measurement of hyperpolarized [1\nobreakdashes-\textsuperscript{13}C]fumarate.
(a) Experimental apparatus.
(b) Schematic of the QuSpin OPM placement around the sample, showing the sensitive axes of the OPMs and the magnetic field lines of a magnetic sample polarized along the $z$ axis.
(c) The hyperpolarization process for [1\nobreakdashes-\textsuperscript{13}C]fumarate. The [1\nobreakdashes-\textsuperscript{13}C]fumarate $J$-couplings are shown in the inset.
(d) Magnetic field `pulse' sequence used for the zero-field experiments.
(e) Zero-field time-domain signal measured along the $x$- and $z$-axes of the OPMs generated from a sample of \textsuperscript{13}C polarized [1\nobreakdashes-\textsuperscript{13}C]fumarate. 
(f) Zero-field spectrum (black) resulting from the Fourier transform of the z-axis signal. A simulated spectrum is shown beneath in magenta.
(g) Magnetic field `pulse' sequence used for the low-field experiments. 
(h) Low-field time-domain signal measured along the $x$- and $z$-axes of the OPMS generated from a sample of \textsuperscript{13}C polarized [1\nobreakdashes-\textsuperscript{13}C]fumarate.
(i) Low-field spectrum (black) resulting from the Fourier transform of the $z$-axis signal. 
A simulated spectrum is shown beneath in magenta.}
\vspace{-14pt}
\end{figure*}

These disadvantages can be circumvented through the use of an alternative modality, zero- to ultralow-field (ZULF) NMR, in which measurements are performed in the absence of a strong magnetic field \cite{Blanchard2021_LtL,JIANG2021ZULF}.
In this field regime, chemical shifts are negligible, and the dominant nuclear-spin interactions are spin-spin ($J$) couplings, which are typically on the order of Hz to hundreds of Hz.
At these low frequencies, conductive materials have a negligible effect on signals, and because susceptibility-induced magnetic-field gradients are proportional to the applied  field strength, resolution is unaffected even in complex heterogeneous samples \cite{Tayler2018nmr,Burueva2020chemical}.
Even though ZULF spectra do not contain chemical-shift information, individual molecules can still be identified with high specificity, since the $J$-couplings provide a means of chemical fingerprinting.\cite{Blanchard2013high,alcicek2021zero} 
This is especially simple for hyperpolarized systems in which only selected chemical species are initially hyperpolarized, and only these molecules and their downstream products contribute to the observable signal.

Nevertheless, at such low frequencies of the electromagnetic signals, inductive detection is inefficient, and ZULF NMR has relied on highly sensitive superconducting quantum interference devices (SQUIDs), and more recently on optically pumped magnetometers (OPMs), that are able to detect magnetic signals with sensitivities on the order of $\text{fT}/\sqrt{\text{Hz}}$.\cite{li2018serf,Tayler2017instrumentation}
The primary use of OPMs is for magnetoencephalography (MEG), which is a separate technique for mapping out brain activity based on the magnetic fields produced by neuronal currents. \cite{Boto2018,KernelFlux}
Despite the exquisite sensitivity offered by OPMs, detection of zero-field NMR signals is approximately 2 to 3 orders of magnitude less sensitive than inductive detection at ca. 7\,T, assuming the same spin polarization and the same sample/detector geometry.
The lower sensitivity of ZULF NMR has so far generally precluded applications, particularly in a biomedical context given the intrinsically low concentrations of biomolecules in biological systems.

In this work we combine recent advances in parahydrogen-based hyperpolarization methods with state-of-the-art magnetometry to demonstrate that metabolic transformations can be observed in zero- and low-field NMR experiments. 
We investigated two model enzymatic processes: the conversion of fumarate into malate, which is used in vivo as a marker of cell necrosis\,\cite{clatworthy2012magnetic} and the conversion of pyruvate into lactate, which is by far the most widely-used metabolic process in hyperpolarization-enhanced imaging.\cite{serrao2016potential} Hyperpolarized ${}^{13}$C-labeled metabolites were prepared via parahydrogen-induced polarization (PHIP)\cite{bowers1987parahydrogen,hovener2018parahydrogen}, which involves chemically reacting a precursor molecule with molecular hydrogen in the nuclear-spin singlet state (parahydrogen) to yield a hyperpolarized product molecule. The entangled state of the $^1$H nuclear-spin pair can be converted into a magnetic state of the ${}^{13}$C spin via a magnetic field sweep.\cite{golman2001parahydrogen,schmidt2022instrumentation} 
Preparing the molecules in a \textsuperscript{13}C-polarized state facilitates the detection of metabolic transformations, since as long as a magnetic field of $\gtrapprox$1\,$\mu$T is applied to the sample, the \textsuperscript{13}C\,nuclei will remain polarized during chemical transformations. We show that the combination of PHIP and ZULF NMR enables the study of biological systems using magnetic resonance in portable devices, without the need for bulky permanent or superconducting magnets.

\begin{figure*}
\centering
\includegraphics[width=0.75\textwidth]{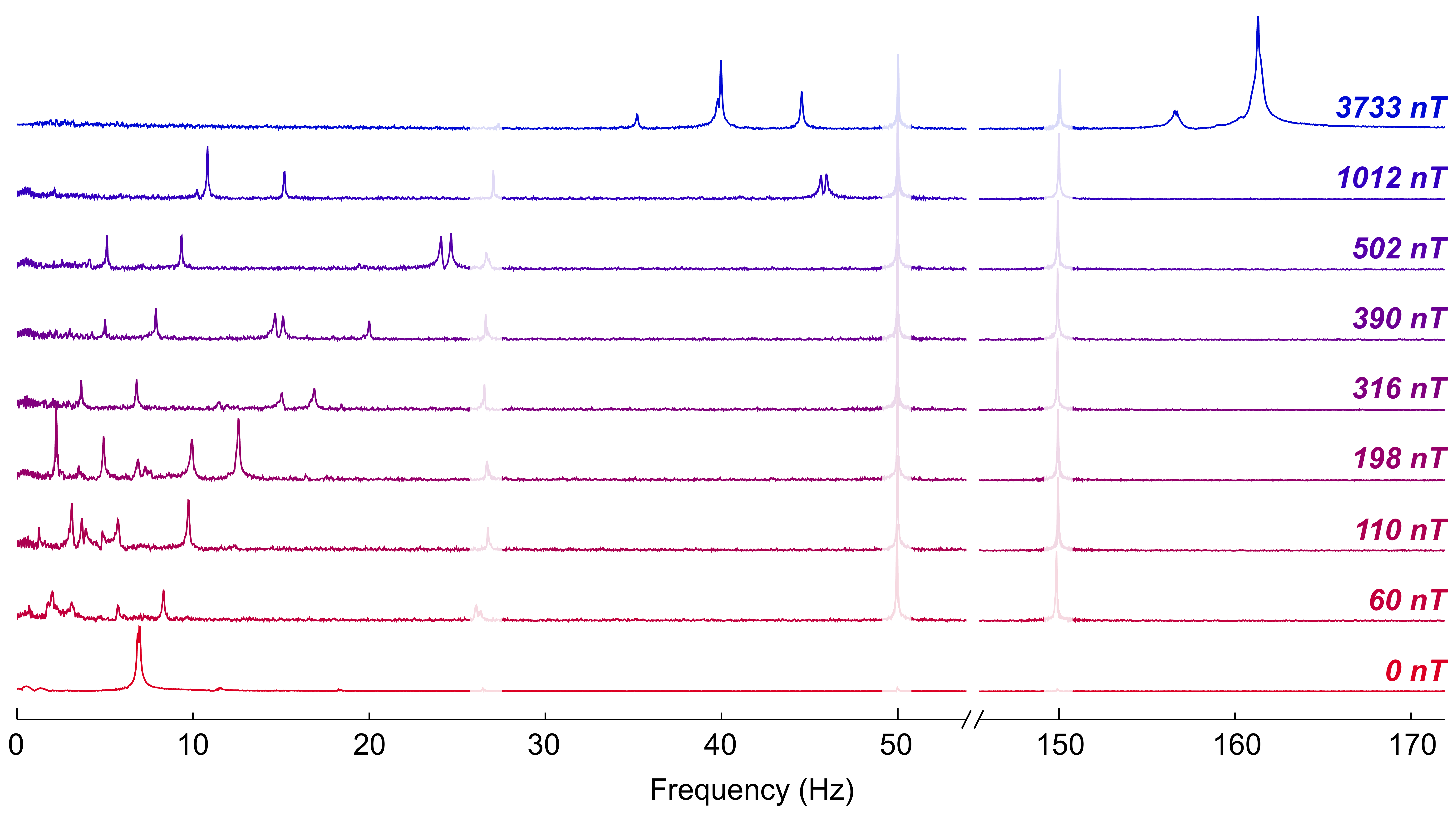}%
\caption{
\label{fig:Variable field} 
Zero- to low-field NMR spectra of [1\nobreakdashes-${}^{13}$C]fumarate, measured using the pulse sequence in Fig.\,\ref{fig:Fig1}(g) (or Fig.\,\ref{fig:Fig1}(d) for the 0\,nT spectrum) at different background magnetic fields. Noise peaks at 27, 50 and 150\,Hz have been partially greyed-out for clarity.
}
\vspace{-14pt}
\end{figure*}

\section{Results}

A schematic of the experimental arrangement is shown in Fig.\,\ref{fig:Fig1}(a).
Samples in 5-mm glass NMR tubes were placed in a magnetic shield and measured with two optically pumped magnetometers.\cite{blanchard2020zero}
A set of three orthogonal Helmholtz coils was present inside the shield to generate magnetic fields for control of the spin states and their dynamics.
Furthermore, the sample was located within a long solenoid coil that pierced through the magnetic shield along the $y$ axis, allowing for a magnetic field to be applied to the sample without significantly affecting the magnetometers.\cite{yashchuk2004hyperpolarized} 
The staggered arrangement of the magnetometers was chosen to reduce common-mode noise while maximizing the signal from the sample; a gradiometric detection scheme.\cite{PhysRevApplied.11.024005} As illustrated in Fig.\,\ref{fig:Fig1}(b), the magnetic field produced by the nuclear spins in the sample has opposite sign at the locations of the two sensors.
In contrast, many noise sources which are centered further from the detectors (e.g., Johnson noise from the magnetic shielding, or vibrations in the room) produce magnetic fields that are more homogeneous at the positions of the OPMs.
By measuring the difference in the outputs from the two sensors, signals from the sample are combined constructively, and common-mode noise is cancelled.
Because the sensors have two sensitive axes, it was possible to independently measure magnetic fields corresponding to magnetization along $x$ and $z$.

Initial experiments to develop and test the pulse sequences were carried out using the hyperpolarized contrast agent [1\nobreakdash-${}^{13}$C]fumarate, since this molecule comprises a simple three-spin system with the magnetic equivalence of the two parahydrogen protons lifted by asymmetric $J$-couplings to the ${}^{13}$C spin.\cite{eills2017singlet} To prepare the molecule, a precursor was hydrogenated using para-enriched hydrogen, and a magnetic field cycle was used to transform the proton singlet order into ${}^{13}$C magnetization.\cite{eills2019real} Using this method we produced solutions of approximately 80\,mM [1\nobreakdash-${}^{13}$C]fumarate at $\approx$\,30\% ${}^{13}$C\ polarization. The hyperpolarization process and molecular $J$-couplings are shown in Fig.\,\ref{fig:Fig1}(c), and further details are provided in the Materials and Methods.

In this work, NMR experiments were performed in two field regimes: zero field, defined as the regime where the nuclear Larmor frequencies are negligible compared to the electron-mediated indirect spin-spin ($J$) coupling, and low field, where the nuclear Larmor frequencies are larger than the $J$-couplings, but chemical shifts are smaller than the resonance width. The pulse sequence for the zero-field NMR experiments is shown in Fig.\,\ref{fig:Fig1}(d).
The magnetization is rotated from the $y$ axis to the $z$ axis, after which the applied magnetic fields are suddenly turned off, resulting in the magnetic signal shown in Fig.\,\ref{fig:Fig1}(e). The spectrum resulting from the Fourier transform of this signal is shown in Fig.\,\ref{fig:Fig1}(f). 
The main feature is the peak at 6.9\,Hz, split by 100\,mHz due to a 2\,nT residual orthogonal field. A simulated spectrum is shown beneath in magenta. All simulations were performed using the SpinDynamica package for Mathematica.\cite{bengs2018spindynamica}

Low-field NMR experiments can be performed by leaving the piercing solenoid on during signal acquisition. As depicted in Fig.\,\ref{fig:Fig1}(g), the magnetization is kept along the $y$ axis by a 20\,$\mu$T field for sample injection, after which the field is reduced to a 3.8\,$\mu$T detection field and the magnetization is rotated to the $x$-axis by a DC pulse along $z$, which results in the precessing magnetic signal shown in Fig.\,\ref{fig:Fig1}(h).
The spectrum resulting from the Fourier transform of this signal is shown in Fig.\,\ref{fig:Fig1}(i).
There are two groups of peaks centered at 40\,Hz and 160\,Hz, corresponding to signals from the \textsuperscript{13}C and \textsuperscript{1}H nuclei of [1\nobreakdash-\textsuperscript{13}C]fumarate, respectively.
Simulated spectra are shown in magenta. The difference in the noise between the magnetic signals in Fig.\,\ref{fig:Fig1}(e) and (h) is predominantly from the 50\,Hz line noise, which is not present in Fig.\,\ref{fig:Fig1}(e) because a low-pass filter was applied to the data during processing.

The zero- and low-field spectra in Fig.\,\ref{fig:Fig1}(f) and (i) are relatively easy to interpret since either the Zeeman interaction can be treated as a perturbation to the $J$-couplings, or vice versa. 
In Fig.\,\ref{fig:Variable field}, we show spectra of [1\nobreakdash-${}^{13}$C]fumarate measured under different background fields, spanning the range between the zero- and low-field regimes. 
The spectra were acquired using the pulse sequence shown in Fig.\,\ref{fig:Fig1}(g) with the exception of the 0\,nT spectrum which was measured using the pulse sequence in Fig.\,\ref{fig:Fig1}(d). 
The magnetic field pulse was set to induce a $\approx4\pi/6$ rotation of protons and $\approx\pi/6$ rotation of ${}^{13}$C nuclei, so that both nuclei were excited.
The spectra exhibit increasing complexity as the field increases and the Zeeman interaction term, which is of the form $B|\gamma_H-\gamma_C|$, approaches the magnitude of the $J$-couplings.\cite{Blanchard2016emagres} 
The number of spectral lines decreases again as the term $B|\gamma_H-\gamma_C|$ begins to dominate, and the spectral lines group into frequency bands corresponding to nuclear spin species. 
It is for this reason that in this work we carried out experiments in the two distinct field regimes, and not between the two.

\begin{figure}
\centering
\includegraphics[width=0.8\columnwidth]{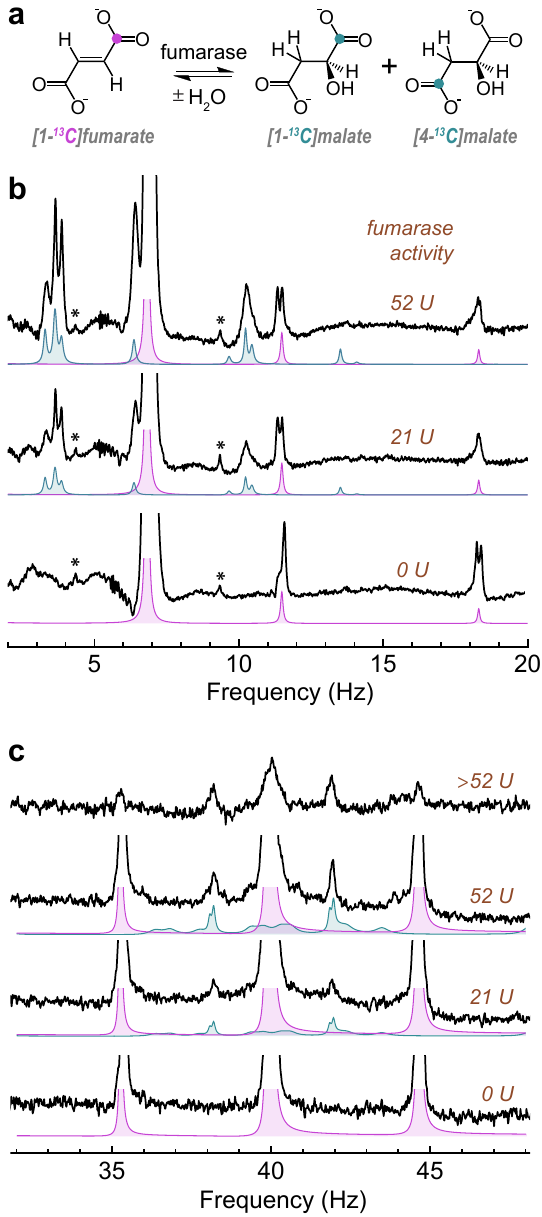}%
\caption{
\label{fig:FumarateToMalate} 
(a) The enzyme-catalysed interconversion between fumarate and malate. The addition of H$_2$O breaks the molecular symmetry, and the 1-${}^{13}$C spin label in fumarate has an equal chance to end up in either the 1- or 4- position in malate. 
(b) Zero-field NMR spectra of the reaction solution. The peaks highlighted with asterisks are carbon satellites from the nat. abun. of [1,4-\textsuperscript{13}C\textsubscript{2}]fumarate molecules.
(c) Low-field NMR spectra at 3.8$\,\mu$T of the reaction solution. 
The amount of fumarase enzyme added to each hyperpolarized fumarate solution prior to detection is shown in brown. 
All spectra are shown with 50\,mHz line broadening. 
Simulations are shown beneath the spectra in pink and teal for fumarate and malate, respectively. 
The simulated spectra are vertically scaled to match the real spectra (see Supporting Information for further details about the simulated spectra, including all spin coupling parameters used).
}
\vspace{-14pt}  
\end{figure}

\subsubsection{Observing fumarate-to-malate conversion}

A reaction solution was prepared with 50\% ${}^{13}$C labelling in the carboxylate position of the acetylene dicarboxylate. 
The same chemical reaction with parahydrogen was used to produce [1\nobreakdash-${}^{13}$C]fumarate, and following the field cycle to polarize the ${}^{13}$C spin, the solution was mixed with phosphate buffer solution containing a varied amount of fumarase to catalyze the reaction shown in Fig.\,\ref{fig:FumarateToMalate}(a). 
The reaction was left to proceed for 10\,s at Earth’s field, after which the solution was measured at zero field using the procedure shown in Fig.\,\ref{fig:Fig1}(d). 
The experiment was repeated three times with differing amounts (0, 21, and 52\,U, where U is enzyme units) of the enzyme, fumarase, to demonstrate how the resulting spectra change as a function of metabolic activity. 
This level of fumarase activity is consistent with reported plasma concentrations following acute kidney injury. \cite{Nielsen2017}
Further experimental details are provided in the Materials and Methods.

The resulting zero-field NMR spectra are shown in Fig.\,\ref{fig:FumarateToMalate}(b), and feature peaks at 6.9, 11.5 and 18.2\,Hz corresponding to [1\nobreakdash-${}^{13}$C]fumarate and peaks at 3.3, 3.6, 3.9, 6.5 and 10.2\,Hz corresponding to [1\nobreakdash-${}^{13}$C] and [4-${}^{13}$C]malate.
The malate peaks are visible for molecules that formed prior to zero-field detection.
The enzymatic reaction does not reach equilibrium on the short timescale of this experiment, and so the malate peaks are largest when 52\,U of enzyme was used, since more malate formed prior to detection.
Simulated spectra for fumarate and malate are shown in pink and teal, respectively, and are generally in agreement with the experimental data (details of the simulations are provided in the Supporting Information). 
The signals are well resolved, even for the partially overlapping signals at 6.5 and 6.9\,Hz. 

Low-field NMR experiments were also performed following the same protocol as above, but using the detection procedure shown in Fig.\,\ref{fig:Fig1}(g).
The resulting ${}^{13}$C spectra are shown in Fig.\,\ref{fig:FumarateToMalate}(c).
The fumarate yields a characteristic 1:2:1 triplet peak pattern centered at 40\,Hz, and malate peaks appear at approximately 38 and 42\,Hz. There are additional malate peaks, visible in the simulations but not the spectra, but these are of significantly lower amplitude. If relaxation effects are neglected, the peaks are all of similar amplitude. To get a better match to the experimental results, dipole-dipole relaxation between the geminal proton pair in malate was included in the simulations. The dipole-dipole relaxation assumed a dipolar coupling of 25\,kHz between the protons, and a rotational correlation time of 50\,ps. This relaxation mechanism was introduced for a 10\,s period prior to the pulse and during signal acquisition.
Importantly, this relaxation mechanism has a negligible effect on the peaks at 38 and 42\,Hz, but they are much lower in amplitude than the malate peaks in the zero-field spectra. We hypothesize there is an additional relaxation mechanism related to hydroxyl proton exchange that is effective at low fields (see Discussion).

\subsubsection{Observing pyruvate-to-lactate conversion}

\begin{figure}
\centering
\includegraphics[width=0.9\columnwidth]{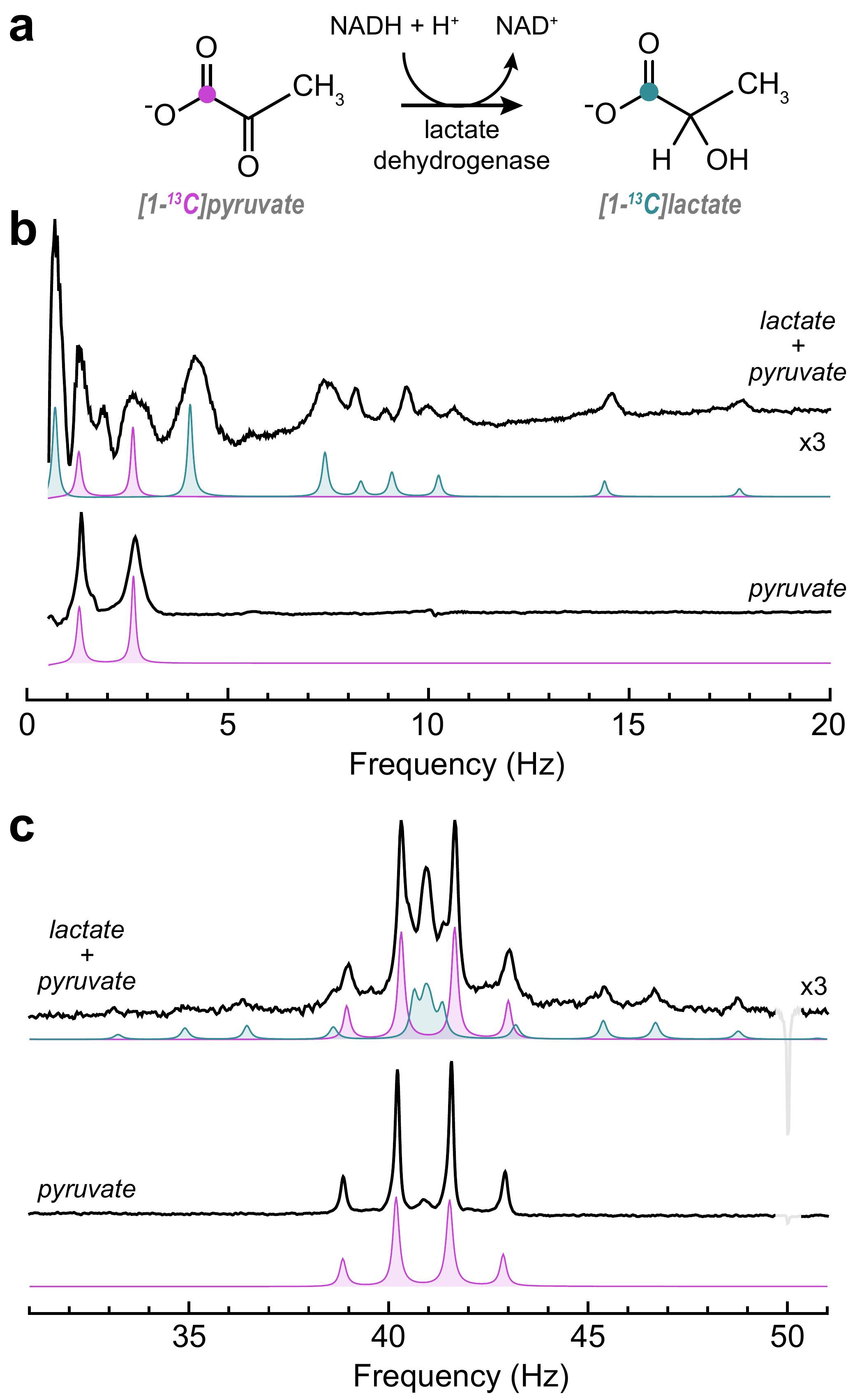}%
\caption{
\label{fig:PyruvateToLactate} 
(a) The enzyme-catalysed conversion of pyruvate to lactate. (b) Zero-field NMR spectra of the reaction solution. (c) Low-field NMR spectra of the reaction solution. For the pyruvate spectra the signal of hyperpolarized pyruvate was acquired directly. For the spectra that show lactate signals, lactate dehydrogenase and an excess of NADH was added to the hyperpolarized pyruvate solution prior to detection. All spectra are shown with 250\,mHz line broadening. Simulations are shown beneath the spectra in pink and teal for pyruvate and lactate, respectively. The simulated spectra are vertically scaled to match the real spectra (see Supporting Information for further details about the simulated spectra, including all spin coupling parameters used).
}
\vspace{-14pt}
\end{figure}

The metabolic conversion of pyruvate to lactate (Fig.\,\ref{fig:PyruvateToLactate}(a)) is by far the most-studied in hyperpolarization-enhanced MRI experiments,\cite{nelson2013metabolic,wang2019hyperpolarized} with over 25 clinical trials currently recruiting or underway around the world.\cite{clinicaltrials}
We formed hyperpolarized [1\nobreakdash-${}^{13}$C]pyruvate via the sidearm hydrogenation method:\cite{reineri2015parahydrogen} propargyl pyruvate was hydrogenated using parahydrogen to produce allyl pyruvate, and a magnetic field cycle was applied to convert the proton polarization into ${}^{13}$C polarization. After this, the [1\nobreakdash-${}^{13}$C]pyruvate was chemically cleaved from the allyl sidearm moiety via base-catalyzed aqueous hydrolysis and then extracted into a phosphate buffer solution to yield a solution of approximately 80\,mM [1\nobreakdash-${}^{13}$C]pyruvate at $\approx$\,3.5\% ${}^{13}$C polarization. Further details are given in the Materials and Methods. This solution was either observed directly, or mixed with a solution containing lactate dehydrogenase and NADH (nicotinamide adenine dinucleotide) leading to metabolic conversion of pyruvate to lactate. The concentration of NADH in the solutions was 150\,mM, an excess compared to the 80\,mM pyruvate, so complete conversion to lactate would be expected if the reaction was given sufficient time to go to completion.

The zero-field results obtained by applying the pulse sequence from Fig.\,\ref{fig:Fig1}(d) are shown in Fig.\,\ref{fig:PyruvateToLactate}(b). The bottom spectrum shows hyperpolarized [1\nobreakdash-${}^{13}$C]pyruvate, which is an XA\textsubscript{3} spin system (one ${}^{13}$C and three methyl protons), which exhibits peaks at $J_\text{AX}$ and $2J_\text{AX}$, where $J_\text{AX}$ is approximately 1.4\,Hz. Above is the spectrum for the sample to which lactate dehydrogenase and an excess of NADH was added, and this shows a number of additional peaks at frequencies up to 18\,Hz. Simulations of the pyruvate and lactate zero-field spectra are shown beneath in pink and teal, respectively.

The low-field results obtained by applying the pulse sequence from Fig.\,\ref{fig:Fig1}(g) are shown in Fig.\,\ref{fig:PyruvateToLactate}(c). Again, the pyruvate spectrum is relatively easy to interpret: the ${}^{13}$C lines are centered at 41\,Hz due to the 3.83\,$\mu$T background magnetic field, and a 1:3:3:1 quartet is observed due to the ${}^{13}$C coupling to the three protons. A small line can be observed at 41\,Hz, which corresponds to a small [1\nobreakdash-${}^{13}$C]parapyruvate impurity.\cite{rios2020pyruvate} The production of lactate can also be seen at low field; a multiplet corresponding to lactate can be resolved in addition to the pyruvate quartet (Fig.\,\ref{fig:PyruvateToLactate}(c)). Simulations of the pyruvate and lactate low-field spectra are shown beneath in pink and teal, respectively.

\subsubsection{Microfluidic implementation}
\begin{figure}
\centering
\includegraphics[width=0.9\columnwidth]{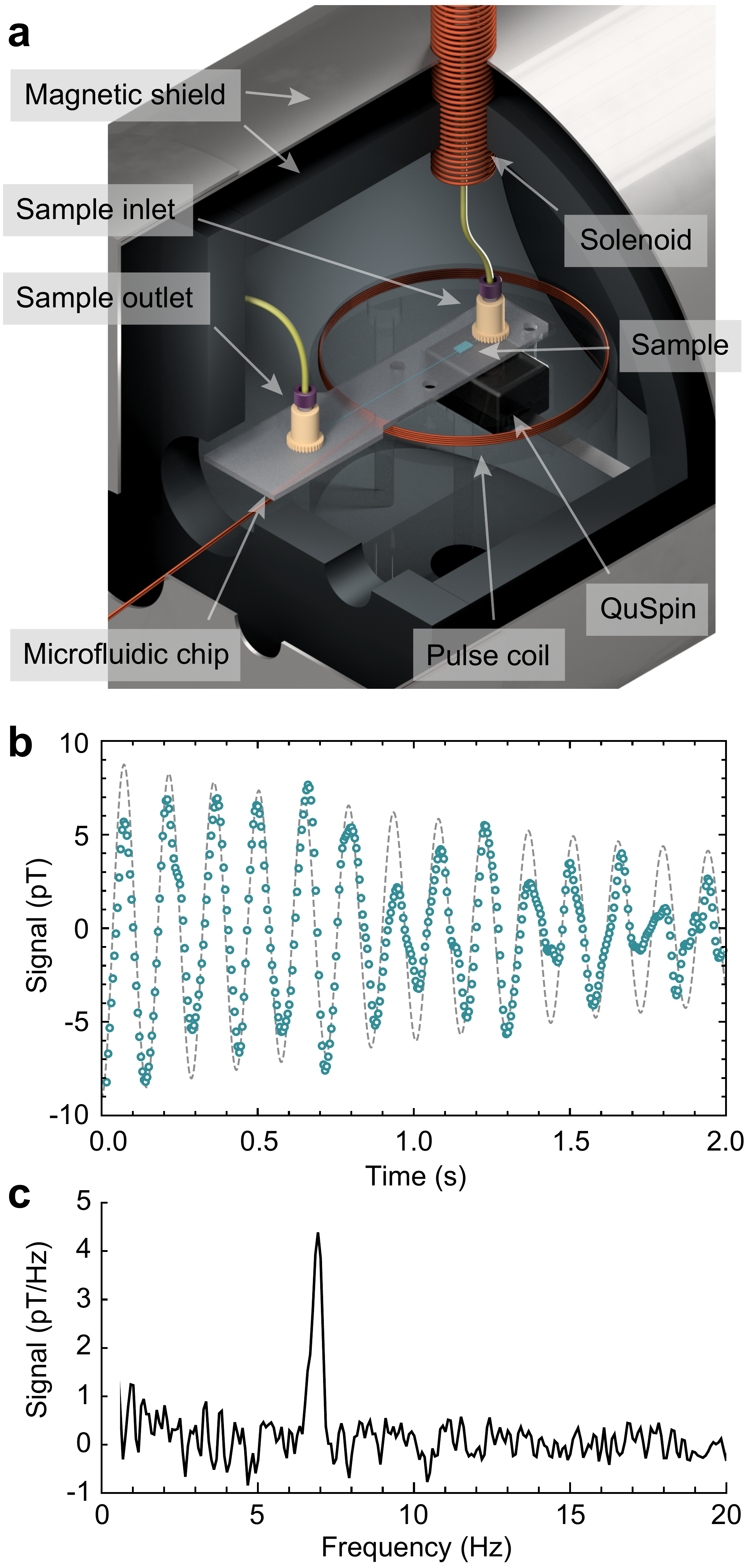}%
\caption{
\label{fig:Microfluidic} 
Measurement of hyperpolarized [1\nobreakdash-\textsuperscript{13}C]fumarate in a microfluidic chip. (a) The experimental setup, showing the microfluidic chip resting on top of a 3D-printed former (made transparent for clarity), which supports a coil and houses the OPM directly beneath the 10\,$\mu$L sample chamber. The PTFE fluid inlet and outlet lines are shown in yellow, and the inlet passes through a guiding solenoid into the shield. (b) The first 2\,s of NMR signal obtained from a sample of [1\nobreakdash-\textsuperscript{13}C]fumarate in the chip (shown by teal circles). A background time signal was acquired by repeating the experiment without a hyperpolarized sample and we show the difference between the real and background signals. A 10\,Hz low-pass filter was applied to remove 50\,Hz line noise, and one in five acquired data points is shown for clarity. A decaying sinusoidal function of the form $S(t)=-9\,\text{cos}(2\pi\,6.95\,t)\,e^{-t/2.5}$ is overlaid on the data in grey. (c) The Fourier transform of the NMR signal without the low-pass filter applied, and with 150\,mHz line broadening.
}
\vspace{-14pt}
\end{figure}

To explore the possibilities to extend this spectroscopic technique into different fields of application, we carried out an experiment with the sample in a microfluidic lab-on-a-chip device, rather than in a 5\,mm NMR tube. The ZULF apparatus was modified to accommodate a polycarbonate microfluidic chip inside the magnetic shield, with an OPM directly under the 10\,$\mu$L sample chamber, and a pulse coil to apply magnetic fields to the sample to excite NMR signals. We carried out the hyperpolarization procedure described above to produce [1\nobreakdash-\textsuperscript{13}C]fumarate, and injected the hyperpolarized solution through a solenoid guiding field into the shield, and into the microfluidic chip. The pulse coil was holding a constant 16\,$\mu$T magnetic field, and after 5\,s this field was switched off and the resulting zero-field NMR signal was detected. A schematic of the apparatus is shown in Fig.\,\ref{fig:Microfluidic}a, and the NMR spectrum obtained from this experiment is shown in Fig.\,\ref{fig:Microfluidic}b. Further experimental details are provided in the Supporting Information.

\section{Discussion}
We have shown that zero- and low-field NMR can be used to measure key steps of metabolic pathways using hyperpolarized molecules. 
The signals from the converted metabolites malate and lactate are lower at elevated (3.8\,$\mu$T) fields compared to zero field, which is not the case for fumarate and pyruvate. 
We attribute this to faster relaxation of malate and lactate at microtesla fields, which may be related to the fact that the rate of hydroxyl exchange is approximately 350\,s$^{-1}$ at pH\,7 and 25$^{\circ}$C.\cite{debrosse2016lactate} 
This is close to nuclear spin Larmor frequencies at microtesla field strengths, and so the fluctuating dipolar field from exchanging protons may act as a relaxation mechanism. 
We performed additional experiments in which we formed [1\nobreakdash-${}^{13}$C]lactate directly via sidearm hydrogenation and allowed it to relax for a fixed time at variable field, followed by measurement of the ${}^{13}$C\ polarization. 
From these experiments we determined that above a field of $\approx$\,30$\mu$T the proposed hydroxyl exchange relaxation mechanism is no longer dominant. 
These results are presented in the Supporting Information.

In the zero-field experiments, the magnetic field sequence to generate the signals was chosen such that the reaction solution remains at relatively high field until signal detection. 
The reason for this is that the hyperpolarization remains as Zeeman order on the \textsuperscript{13}C nucleus during the metabolic transformation so the change in $J$-coupling topology does not significantly affect the hyperpolarized spin order. 
We expect that if the metabolic transformations occurred at zero field, the change in the $J$-coupling Hamiltonian, which occurs at different time points for each molecule, would cause loss of coherence and hence significant polarization losses.\cite{barskiy2019zero} 
It is only immediately before signal acquisition that the magnetic field is switched to zero, and we observe malate/lactate that formed prior to this, i.e., we do not see the molecules that form during signal acquisition.

Generally, high-field magnetic resonance imaging of metabolism using hyperpolarized biomolecules is performed using a train of small-angle pulses, so as not to significantly perturb the hyperpolarized spin order in each experiment. 
This allows one to collect a time-series of spectra to study metabolic flux, and extract information about the reaction kinetics. 
By contrast, in the experiments reported here, the magnetic field sequence converts the hyperpolarized spin order into observable signals which are detected once; in this sense it is a ‘single-shot’ experiment. 
An important next step will be to implement reaction monitoring of the hyperpolarization-enhanced signals using a magnetic field sequence that allows for weak perturbation of the hyperpolarized spin bath many times, to collect a series of spectra. 
The only limitation for low-field experiments would be signal-to-noise, which is lower when using small flip-angle pulses. 
For the zero-field experiments it is more challenging, since molecular transformations occurring at zero field change the $J$-coupling network nonadiabatically and likely lead to irretrievable signal loss. 
We note that reaction monitoring was previously demonstrated with zero-field NMR,\cite{Burueva2020chemical} but in that experiment the hyperpolarization was continually refreshed by the ongoing reaction, which is not the case for most biologically-relevant reactions involving hyperpolarized metabolites.

The presence of quadrupolar nuclei at zero field can pose problems. At zero field, nuclear spins with mutual $J$-coupling become strongly coupled, since the Zeeman energy differences are absent and the $J$-coupling between them dominates. 
Consider two spins with a difference in Zeeman polarization that become strongly coupled in a nonadiabatic manner (e.g., by quickly turning off an external magnetic field, or by chemical reaction); the polarization of the two spins will begin to exchange in an oscillatory manner, with the frequency given by the $J$-coupling. This oscillating magnetization is the observable in many zero-field NMR experiments. 
However, this makes the hyperpolarized spin order particularly sensitive to the presence of $J$-coupled quadrupolar nuclei (e.g., \textsuperscript{2}H or \textsuperscript{14}N), since polarization that oscillates onto these spins can rapidly relax.\cite{barskiy2020zero} 
In fumarate experiments, the reaction to generate fumarate is carried out in D\textsubscript{2}O to prolong the hyperpolarization lifetime, but the conversion to malate should not be carried out in pure D\textsubscript{2}O since all malate molecules would then contain deuterium.\footnote{We performed separate experiments in which the enzyme reaction was carried out in pure D\textsubscript{2}O, and observed fumarate signals but no NMR signals from malate.}
For this reason we used a 1:1 H\textsubscript{2}O/D\textsubscript{2}O mixture as the solvent for the metabolic reaction, so approximately 50\% of the malate molecules are protonated rather than deuterated. 
Novel methods now enable decoupling of deuterium nuclei, which should help to alleviate this problem by suppressing the polarization exchange.\cite{bodenstedt2021decoupling,dagys2022deuteron} 
Regardless, for in vivo experiments the fumarate would first be purified and then brought into an H\textsubscript{2}O solution using established methods.\cite{knecht2021rapid}

The gradiometric detection scheme used in this work is shown schematically in Fig.\,\ref{fig:Fig1}(b). 
Taking the sample to be a long cylinder homogeneously magnetized transverse to the cylindrical axis, the resulting magnetic field takes a two-dimensional dipole shape, such that the field magnitude is constant at a given radius from the center of the sample.
Hence, in these experiments, the signal was acquired simultaneously in both the $x$- and $z$-channels of two QuSpin OPMs, and during processing the signal difference was taken for each axis. 
This gradiometric detection scheme has two advantages: (1) the observable signal is enhanced by detecting with multiple sensors (up to a factor of 2), and; (2) acquiring the differential signal is an effective way to cancel common-mode noise (magnetic noise that is the same at the positions of both OPMs).

In this work the low-field experiments were carried out at relatively low frequency ($<$200\,Hz) since the bandwidth of the magnetometers here was 135\,Hz. This is certainly not a fundamental limitation of the commercial OPMs used in this work, which have been shown to be capable of operation at up to 1.7\,kHz.\cite{Savukov2017QuSpin}

Most experiments were carried out with 5\,mm NMR tubes, with a sample volume of $\approx$600\,$\mu$L, and a detected volume of approximately 100-200\,$\mu$L. In the microfluidic experiment we were able to observe a 10\,$\mu$L sample in a single scan, albeit with a reduced signal-to-noise ratio. A quantitative comparison between the signal in macro and microfluidic experiments is challenging due to the different experimental procedures that were employed to obtain the spectra. Although the microfluidic chip used did not involve fluidic complexity, this is a strength of lab-on-a-chip devices, and indeed it has been shown that the chemical reaction to hyperpolarize fumarate can be carried out on a chip\cite{barker2022direct}. Microfluidic NMR is hindered in widespread use in part due to the need for specially-designed probes for high-field implementation, as well as limited resolution due to magnetic field gradients at the interface between the chip and sample; ZULF NMR does not suffer from these limitations, and might prove to be a useful spectroscopic technique in this field.

\section{Conclusions}
In conclusion, we have shown that biochemical reactions can be quantified by NMR at zero magnetic field. 
We used a gradiometer based on optically-pumped magnetometers to observe the conversion of [1\nobreakdash-\textsuperscript{13}C]fumarate to [1\nobreakdash-\textsuperscript{13}C]malate and [4-\textsuperscript{13}C]malate, and [1\nobreakdash-\textsuperscript{13}C]pyruvate to [1\nobreakdash-\textsuperscript{13}C]lactate at both zero and low field. 
We also report zero-field NMR spectroscopy of hyperpolarized [1\nobreakdash-\textsuperscript{13}C]fumarate in a microfluidic chip. To our knowledge, this is the first demonstration of zero-field NMR in a microfluidic system. Together with the ability to follow biochemical processes, this opens up interesting possible applications in the context of lab-on-a-chip cell culture devices.

The particular appeal of OPM detection is that the advanced magnetometer arrays and image reconstruction algorithms that have been made available for MEG \cite{KernelFlux} could be used directly for low-field metabolic MRI.
This would, for example, enable metabolic imaging of patients with medical implants, free from issues with susceptibility and rf shielding effects.
We envision a combined MEG/MRI modality capable of simultaneously measuring electromagnetic neuronal activity, brain metabolism, and correlations between the two. 
The step to in vivo applications will naturally be challenging as a result of sensitivity reductions, due to the greater distance between the sensor and reactions occurring inside the body, and also the lower concentration of biomolecules in vivo. 
On the other hand, the ever-improving sensitivity of magnetometers, developments in pulse excitation schemes, and increasing polarization levels that can be generated on hyperpolarized biomolecules have consistently improved NMR sensitivity over the last 70 years, and will likely continue to do so.

\bibliography{MetabolicZULF}

\begin{acknowledgement}
This project has received funding from the European Union’s Horizon 2020 Research and Innovation Programme under the Marie Sk\l{}odowska-Curie Grant Agreement 766402 and the FET-Open AlternativesToGd Proposal n. 858149. as well as by the DFG (Project ID 465084791). This work has been supported by an EPSRC iCASE studentship EP/R513325/1 to SJB, co-funded by Bruker UK Ltd.
\end{acknowledgement} 

\newpage

\section*{Supplementary materials}
Materials and Methods\\
Spectral simulations\\
Hyperpolarization decay at low field\\

\section{Supplementary Materials}

\subsection{Materials and Methods}

All high-field NMR experiments (unless otherwise stated) were performed in a 1.4\,T \textsuperscript{1}H-\textsuperscript{13}C dual resonance SpinSolve NMR system (Magritek, Aachen, Germany).

Parahydrogen at $>$95\% enrichment was generated by passing hydrogen gas ($>$99.999\% purity) over a hydrated iron oxide catalyst in a cryostat operating at 30\,K (Advanced Research Systems, Macungie, U.S.A.).

For magnetic field sweeps for polarization transfer, an MS-2 four-layer mu-metal magnetic shield (Twinleaf LLC, Princeton, U.S.A.) was used to provide a $10^4$ shielding factor against external magnetic fields. No static shim fields were required, since the residual field within the shield was on the order of 5\,nT. The time-dependent applied magnetic field was generated using the built-in $B_y$ shim coil, with current supplied with an NI-9263 analog output card (National Instruments, Aachen, Germany) with 10 µs time precision. The guiding magnetic field for sample transport in and out of the magnetic shield was provided with a handmade solenoid, with current supplied with a Keysight U8001A current source (Keysight Technologies, Böblingen, Germany).

\textbf{\textit{Fumarate Experiments}}

The precursor solution for all fumarate experiments was 250\,mM acetylene dicarboxylic acid monopotassium salt, 250\,mM sodium sulphite and 7\,mM ruthenium catalyst [RuCp*(CH\textsubscript{3}CN)\textsubscript{3}]PF\textsubscript{6} in D\textsubscript{2}O. 250\,mM NaOD was added to the solution to be equimolar with the starting material. All chemicals were purchased from Sigma Aldrich.

The reactor for fumarate experiments was constructed from stainless steel with an internal volume of 20\,mL. PEEK (polyether ether ketone) tubing of 1/8\,in O.D. was used to flow the para-enriched hydrogen gas, and 1/16\,in O.D. 0.5\,mm I.D. PTFE (polytetrafluoroethylene) capillaries were used for all solution flow. These tubes were connected to the reactor via 1/4-28 PEEK fittings (part numbers P-249 and P-349, IDEX LLC, Oak Harbor, U.S.A.), and to the gas flow-control manifold via Swagelok fittings (Swagelok, Frankfurt, Germany). The reactor was wrapped in two heater mats (part number 798-3753, RS Components, Corby, U.K.), and these were connected to a power supply, with the current set to heat the reactor to 85$^{\circ}$C.

For fumarate hydrogenation experiments, a 1.5\,mL aliquot of precursor sample was loaded into the reactor via syringe injection, and given 10\,s to reach 85\,$^{\circ}$C. The reactor was sealed, and parahydrogen was bubbled into the reaction solution at approximately 6\,L/min for 60\,s at a pressure of 8.5\,bar. The flow rate was set by a needle valve at the outlet of the gas manifold. After bubbling, the sample was pneumatically ejected through a PTFE capillary into a 10\,mm NMR tube in the magnetic shield underneath by manually opening a two-way microfluidic valve. To prevent the sample from passing through any fields that could lead to undesired state-mixing during sample transport in/out of the shield, a piercing solenoid was used to provide a 10\,$\mu$T guiding field. Upon landing in the 10\,mm NMR tube, the field was nonadiabatically (rapidly) dropped to 50\,nT and then adiabatically increased to 1\,$\mu$T in 2\,s to transfer the proton singlet order into \textsuperscript{13}C magnetization. The solution was extracted through a 1/16\,in PTFE capillary into a syringe.

For the zero-field experiments to observe fumarate (without conversion to malate), the hyperpolarized fumarate sample was syringe-injected into a 5\,mm NMR tube, and this was dropped into the zero-field spectrometer for signal acquisition.

For the experiments to observe enzymatic conversion to malate, 150\,$\mu$L of the hyperpolarized fumarate sample was injected into a 5\,mm NMR tube containing a stated amount of fumarase enzyme in 450\,$\mu$L 500\,mM phosphate buffer solution at pH\,7. The NMR tube was shaken for 5\,s and then dropped into the zero-field spectrometer for signal acquisition.

\textbf{\textit{Pyruvate Experiments}}

The precursor solution for all pyruvate experiments was 400\,mM propargyl pyruvate and 20\,mM rhodium catalyst [Rh(dppb)(COD)]BF\textsubscript{4} in a solvent of 95:5 (v/v) CDCl\textsubscript{3}:ethanol-d\textsubscript{6}, which was prepared by dissolving the catalyst in the solvent, and adding the propargyl pyruvate immediately (tens of seconds) before each experiment. The propargyl pyruvate precursor was synthesized in-house, and all other chemicals were purchased from Sigma Aldrich.

For pyruvate hydrogenation experiments, a 250\,$\mu$L aliquot of precursor sample was loaded into a 5\,mm low pressure/vacuum NMR tube via syringe injection and the tube was pressurized to 7\,bar pH\textsubscript{2} pressure. The sample was heated to 120\,$^{\circ}$C with a heatgun, and then the tube was vigorously shaken for 4\,s while under pressure. After shaking, the whole NMR tube was placed in the magnetic shield for a 4\,s magnetic field sweep. After the field sweep, the solution was ejected from the NMR tube using a valve at the top and the internal gas pressure to push the solution through a 1/16\,in PTFE capillary into a 10\,mm NMR tube containing 300\,$\mu$L 400\,mM NaOH at 80\,$^{\circ}$C for the hydrolysis. The resulting mixture was allowed to settle for 2\,s and the aqueous layer was extracted through a 1/16\,in PTFE capillary into a syringe containing 300\,$\mu$L acidified 500\,mM phosphate buffer solution to neutralize the pyruvate solution at pH\,7.

For the zero- and low-field experiments, this solution was injected into a 5\,mm NMR tube, and this was dropped into the zero-field spectrometer for signal acquisition.

For the experiments to observe conversion to lactate, the syringe used to extract the aqueous phase from the 10\,mm NMR tube contained 200\,$\mu$L of acidified phosphate buffer solution, such that, after mixing, the solution ended at pH\,7. Another 100\,$\mu$L of phosphate buffer solution contained 50\,mg NADH and 25\,$\mu$L lactate dehydrogenase (Sigma Aldrich, L7525), and this solution was held in the 5\,mm NMR tube. The pyruvate solution was injected into the 5\,mm NMR tube and mixed with the enzyme solution by vigorous shaking for 5\,s, before being dropped into the zero-field spectrometer for signal acquisition.

A modified experimental procedure from the established protocol reported in Ref.\,\cite{cavallari2020vitro} was adopted to increase the volume of the solution and maximize pyruvate concentration (and hence signal) for this proof-of-concept demonstration. For in vivo/in vitro studies, a protocol has to be established in order to attain a biologically compatible solution of the substrate at high concentration.

\textbf{\textit{Microfluidic experiment}}
The microfluidic experiment was carried out in an MS-1LF (Twinleaf LLC, Princeton, U.S.A.) shield. The coil consisted of 10 turns of 0.5\,mm O.D. copper wire wound around a former, with a diameter of 35\,mm. A QuSpin was held inside and the chip was screwed in place above, with the sample chamber directly over the sensitive region of the OPM. The chip itself was made of three layers of polycarbonate, 0.25\,mm thick for the top and bottom layers and 1\,mm thick for the middle layer. The layers were bonded using a plasticiser (5\% v/v  of dibuthyl phalete in isopropyl alcohol) under heat and pressure (85$^{\circ}$C, 5\,tonnes) for 20\,minutes. The chip had inlet+outlet ports connected to 1/16\,in O.D. PTFE tubing via PEEK nanoports (Kinesis GmbH, Germany). The sample chamber had dimensions of 4\,$\times$\,2.5\,$\times$\,1\,mm, and hence a volume of 10\,$\mu$L.
A hyperpolarized sample of $\approx$80\,mM [1\nobreakdash-\textsuperscript{13}C]fumarate in D\textsubscript{2}O was prepared as described above (at $\approx$30\% \textsuperscript{13}C\,polarization), and syringe-injected through the PTFE tubing to fill the microfluidic chip. The pressure was released to allow the fluid to settle for 3\,s, after which the solenoid and pulse coil were switched off non-adiabatically and the signal acquired for 4\,s.

\textbf{\textit{Equipment for zero-field NMR experiments}}

For the zero-field experiments, an MS-1LF four-layer magnetic shield (Twinleaf LLC, Princeton, U.S.A.) was used to provide a $10^4$ shielding factor against external magnetic fields. To provide a stable current for 3-axis shimming using the MS-1LF built-in $B_x$, $B_y$ and $B_z$, shim coils, a CSB-20 3-output current supply (Twinleaf LLC, Princeton, U.S.A.) was used, which has current noise on the order of 100\,pA\,Hz$^{-1/2}$. A hand-wound piercing solenoid produced a guiding magnetic field for sample transport in and out of the magnetic shield, and provided a stable field at the sample position for low-field experiments. The $B_y$ and $B_z$ Helmholtz coils were hand-wound around a 3D-printed coil former, with 10 turns, and 40\,mm coil radius. The time-dependent Helmholtz fields were produced by a NI-9263 analog output card, and AE Techron 7228 amplifiers (AE Techron Elkhart, IN 46516 U.S.A.) were used to amplify the current.

For signal detection, two QuSpin optically-pumped magnetometers were held in place in the 3D-printed coil former, oriented as shown in Fig.\,\ref{fig:Fig1}(a). The distance between the center of the sample and the center of each magnetometer cell was approximately 13 mm. The analog outputs of the QuSpins were read out by a National Instruments NI 9215 analog input card at 50000\,samples/s, downsampled to 5000\,samples/s.

\subsection{Spectral simulations}
\begin{figure}
\centering
\includegraphics[width=\columnwidth]{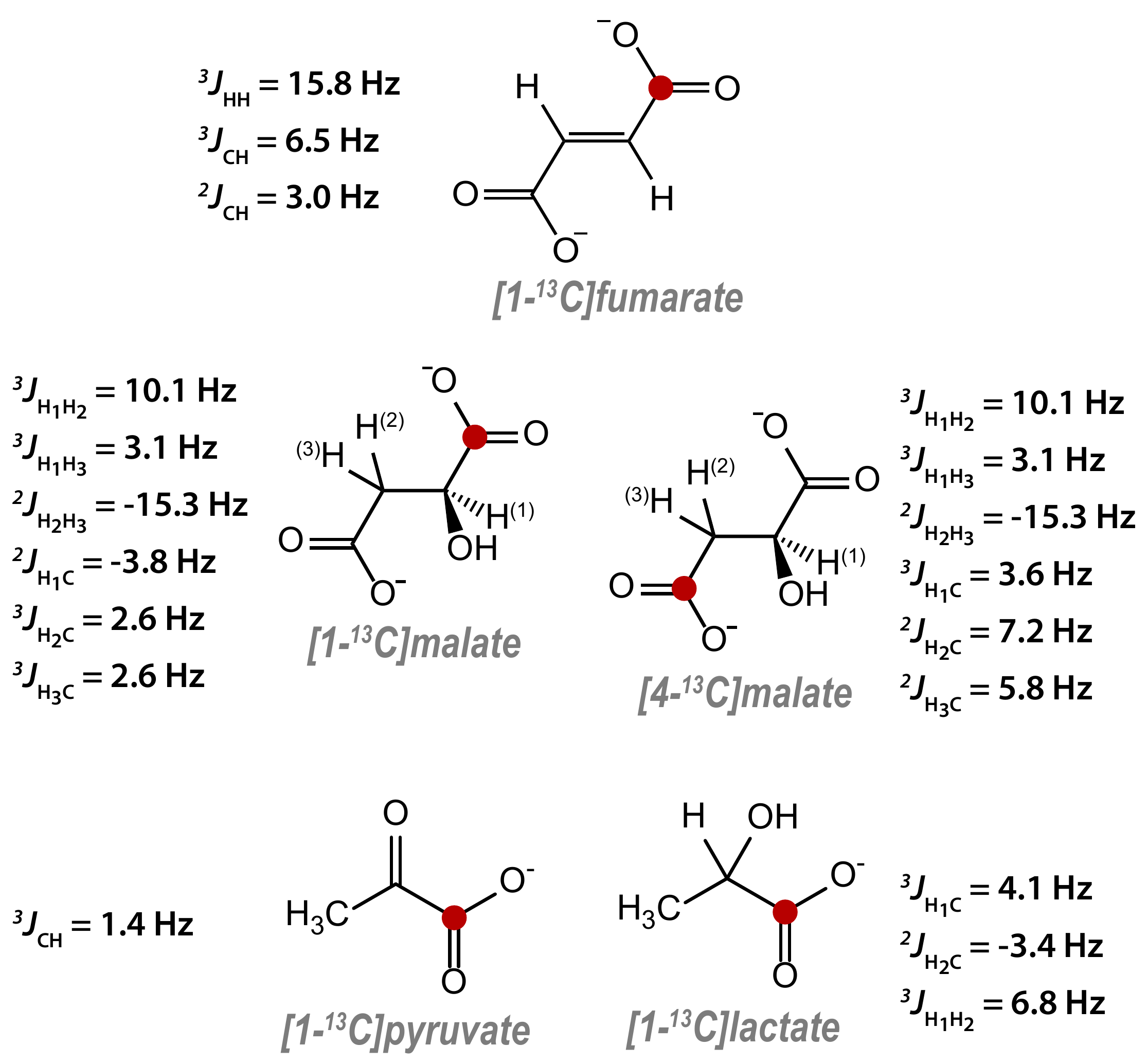}%
\caption{
\label{fig:J-couplings} 
The $J$-couplings used for all simulations.}
\vspace{-14pt}
\end{figure}

The simulations of NMR spectra were carried out using the SpinDynamica package for Mathematica,\cite{bengs2018spindynamica} and the $J$-coupling constants shown in Fig.\,\ref{fig:J-couplings}. Initial guesses of the $J$-couplings were determined by high-field NMR, and these were modified by $<1$\,Hz each in all cases, to fit the simulations to the spectra.
Simulations involved propagating an initial density operator through the pulse sequences shown, and detecting the magnetic signal produced by the sample, which corresponds to the operator \[ \sum_{i} \hbar \gamma_\text{i} I^i_z\text{,} \]
where $\gamma_\text{i}$ is the gyromagnetic ratio of spin $i$, and $I^i_z$ is the $z$ angular momentum operator for the $i$-th spin.

For all simulations relaxation was included phenomenologically (with an equal time constant for all spins) during detection to match the peak linewidths to the experimental data. For the low-field simulations of [1\nobreakdash-\textsuperscript{13}C]- and [4-\textsuperscript{13}C]malate, dipole-dipole relaxation between the geminal proton pair was also included in order to get a better match between simulation and experimental results. The dipole-dipole relaxation assumed a dipolar coupling of 25\,kHz between the protons, and a rotational correlation time of 50\,ps. This relaxation mechanism was introduced for a 10\,s period prior to the pulse, and during signal acquisition.

\subsection{Hyperpolarization decay at low field}
To better-understand the proposed low-field relaxation of [1-\textsuperscript{13}C]lactate, we carried out experiments in which we hyperpolarized [1-\textsuperscript{13}C]lactate via side-arm hydrogenation directly, and allowed it to relax for 5\,s at variable field prior to signal acquisition. We did the same for [1-\textsuperscript{13}C]pyruvate so the results could be compared.

Hyperpolarized [1\nobreakdash-\textsuperscript{13}C]lactate and [1\nobreakdash-\textsuperscript{13}C]pyruvate were obtained according to the procedure already reported by Cavallari et al.\cite{Cavallari2017,Cavallari2018}.
A solution of approximately 250\,mM of the propargyl ester (propargyl-[1\nobreakdash-\textsuperscript{13}C]lactate or propargyl-[1\nobreakdash-\textsuperscript{13}C]pyruvate) and 13.8\,mM of the catalyst [Rh(cod)dppb][BF\textsubscript{4}] in 100\,$\mu$L of 95:5 (v/v) CDCl\textsubscript{3}:ethanol, was loaded into pressurizable NMR sample tubes, and these were pressurized with parahydrogen gas (86$\%$ enriched, to a pressure of 2\,bar at liquid nitrogen temperature). 
The tubes were kept frozen (77\,K) to prevent any chemical reaction, until the start of the hyperpolarization experiment. 
To initiate the hydrogenation reaction, the NMR tube was heated in a hot water bath at 80\,$^{\circ}$C for 7\,s, shaken vigorously for 3\,s, then opened to release the parahydrogen pressure. The tube was then placed in a mu-metal shield (Bartington TLMS-C200) equipped with a coaxial solenoid, through which a magnetic field cycle (MFC) was applied, to obtain the spin-order transfer from the parahydrogen protons to \textsuperscript{13}C. The magnetic field profile applied in all the experiments consists of a diabatic passage from 1.5\,$\mu$T (initial magnetic field) to 50\,nT, and then up to 10\,$\mu$T in 4\,s with an exponential profile.

To hydrolyse the \textsuperscript{13}C-polarized ester (allyl-[1\nobreakdash-\textsuperscript{13}C]-pyruvate/lactate) a hot 260\,$\mu$L solution of 230\,mM NaOH and 50\,mM ascorbate was injected with 1.5\,bar argon pressure. Hydrolysis of the ester occured in a few seconds and the aqueous phase containing the hydrolysed pyruvate/lactate was extracted, whereas the catalyst was retained in the organic phase. An acidic buffer (135\,mM HEPES + 60\,mM HCl) was added to the aqueous phase to obtain a neutral pH (7.4).

\begin{figure}
\centering
\includegraphics[width=1\columnwidth]{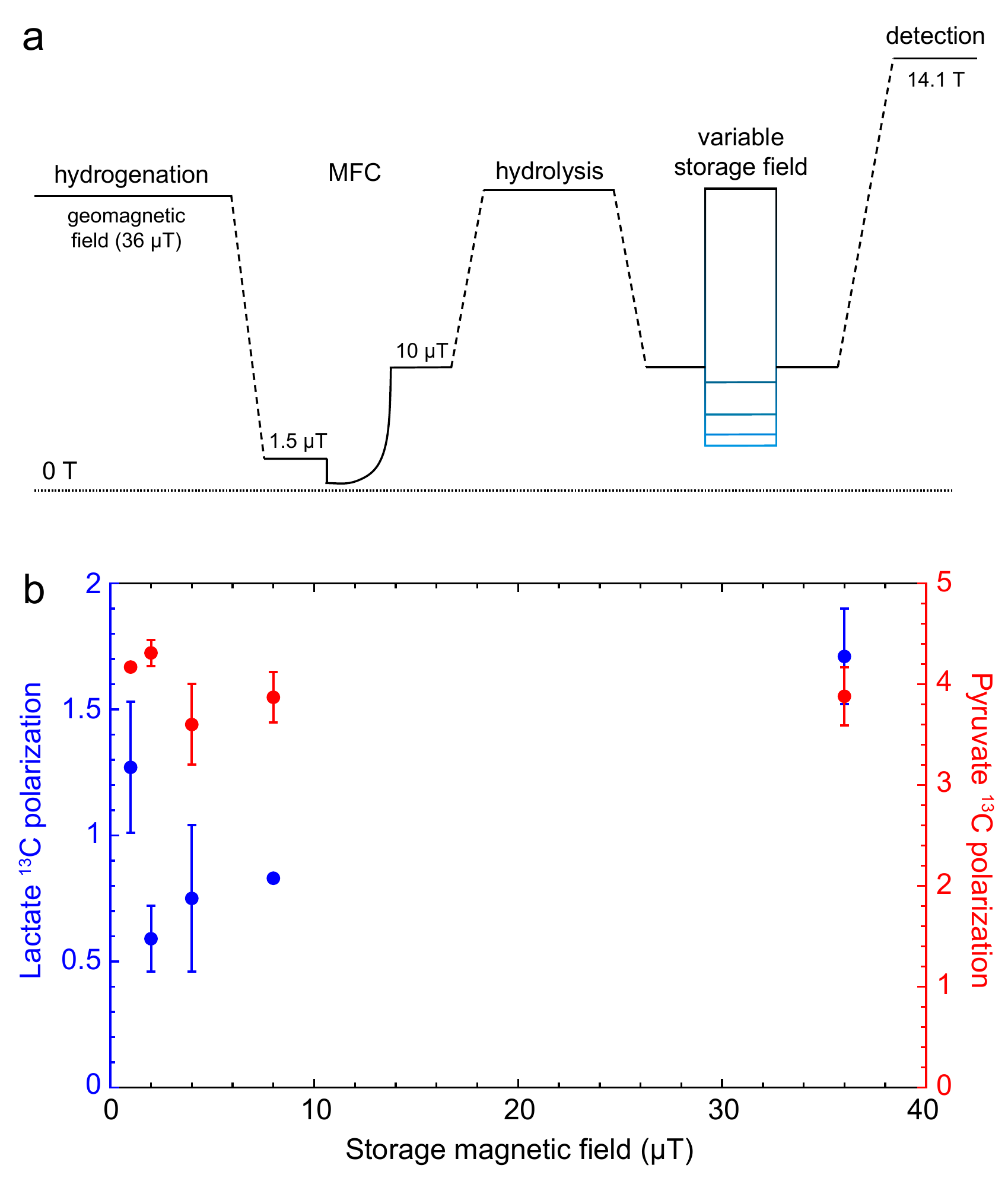}%
\caption{
\label{fig:EC} 
(a) External magnetic field strength during the experiment, with the experimental procedure labelled. (b) Polarization level of [1\nobreakdash-\textsuperscript{13}C]pyruvate and [1\nobreakdash-\textsuperscript{13}C]lactate after storage at various fields for 5\,s before signal acquisition.
}
\vspace{-14pt}
\end{figure}

To investigate the effect of the external magnetic field on the relaxation of [1\nobreakdash-\textsuperscript{13}C]lactate and [1\nobreakdash-\textsuperscript{13}C]pyruvate, the hyperpolarized solutions were held at different magnetic fields (1, 2, 4, 8, and 36\,$\mu$T) for 5\,s before acquisition of the \textsuperscript{13}C NMR spectrum at high field using a 14.1\,T NMR Bruker Avance spectrometer. The protocol is shown in Fig.~\ref{fig:EC}a. The magnetic fields were generated in the $\mu$-metal shield by controlling the current through the solenoid coil, except 36\,$\mu$T which was the laboratory magnetic field.
Each measurement was performed using a freshly-prepared hyperpolarized sample. The results are shown in Fig.~\ref{fig:EC}b. There is a reduction in lactate signal after storage at 2, 4, and 8\,$\mu$T field, but this reduction is less pronounced at 1\,$\mu$T and 36\,$\mu$T. By comparison, the pyruvate signal does not show this field dependence, and so we posit that the enhanced lactate relaxation is due to hydroxyl proton exchange. This question is subject to ongoing investigation.

\end{document}